\documentclass[aps,prb,amsmath,amssymb,preprint]{revtex4-1}
\usepackage{float}
\usepackage{times}
\usepackage{graphicx}
\usepackage{graphics}
\usepackage{color}

\begin{document}
\title{Configuration dependent demagnetizing field in assemblies of interacting magnetic particles}
\author{J. M. Mart\'{i}nez-Huerta}
\affiliation{Instituto de F\'{i}sica, Universidad Aut\'{o}noma de San Luis Potos\'{i}, Av. Manuel Nava 6, Zona Universitaria, 78290 San Luis Potos\'{i}, SLP, M\'{e}xico}
\author{J. De La Torre Medina}
\affiliation{Facultad de Ciencias F\'{i}sico Matem\'{a}ticas, Universidad Michoacana de San Nicol\'{a}s de Hidalgo, Avenida Francisco J. M\'{u}jica S/N, 58060, Morelia, Michoac\'{a}n, Mexico}
\author{L. Piraux}
\affiliation{Institute of Condensed Matter and Nanosciences, Universit\'{e} Catholique de Louvain, Place Croix du Sud 1, B-1348, Louvain-la-Neuve, Belgium}
\author{A. Encinas}
\email[Contact Author; electronic mail: armando.encinas@gmail.com]{}
\affiliation{Instituto de F\'{i}sica, Universidad Aut\'{o}noma de San Luis Potos\'{i}, Av. Manuel Nava 6, Zona Universitaria, 78290 San Luis Potos\'{i}, SLP, M\'{e}xico}

\date{\today}

\begin{abstract}
\normalsize{A mean field model is presented for the configuration dependent effective demagnetizing and anisotropy fields in assemblies of exchange decoupled magnetic particles of arbitrary shape which are expressed in terms of the demagnetizing factors of the particles and the volumetric shape containing the assembly. Perpendicularly magnetized 2D assemblies have been considered, for which it is shown that the demagnetizing field is lower than the continuous thin film. As an example of these 2D systems, arrays of bistable cylindrical nanowires have been characterized by remanence curves as well as ferromagnetic resonance, which have served to show the correspondence of these measurements with the model and also to validate the mean field approach. Linear chains of cylinders and spheres have been analyzed leading to simple expressions to describe the easy axis rotation induced by the interaction field in chains of low aspect ratio cylindrical particles, and the dipolar magnetic anisotropy observed in the linear chain of spheres. These examples serve to underline the dependence on the dipolar interaction field and effective demagnetizing factor on the contributions that arise from the shape of the outer volume.}

\end{abstract}
%
\maketitle

Assemblies or arrays of exchange decoupled magnets represent a very large class of magnetic systems which are of great interest in different subjects and problems in magnetism at both, fundamental and technological level.\cite{1,2,3,4}
 These include assemblies in which the particles are arranged or dispersed on a substrate or embedded in different media such as liquids, polymers or other non magnetic materials.\cite{5,6,7} Depending on their spatial disposition, and thus the shape or outer volume that contains the assembly, they form one, two or 3D heterostructures or composites.\cite{5,6,terris2,lau,9}
 
The magnetic properties of a single isolated particle are mainly governed by the shape and magnetocrystalline anisotropy and eventually a magnetoelastic contribution. In an assembly of many of such particles, their individual properties are further modified by the interaction field produced by the other particles. If the particles are not allowed to come into contact, then the interaction is purely dipolar and has the same magnetostatic origin as the demagnetizing field and shape anisotropy associated to each particle. In this sense, from a mean field perspective, an effective demagnetizing field which includes the particle self demagnetizing field as well as the dipolar interaction is a common feature of any assembly of exchange decoupled assembly of particles.\cite{richter1,coey-book} Consequently, these considerations should apply to any assembly of particles, whether the particles are magnetically hard or soft, and regardless of the approach or level of approximation used to describe them. In particular, it should be possible to interpret the vast phenomenology of effects associated to the dipolar interaction or the effective demagnetizing field, within this framework.\cite{supermagnetic,cowburn1,guslienko1,guslienko2,wires,Encinas,adeyeye11}  However, this is usually complicated by the difficulty of finding either adequate expressions or reliable measurements for the dipolar interaction. Furthermore, extensions to include the magnetization or configuration dependency of the dipolar interaction field adds additional difficulties for establishing a suitable description for these systems.

A coherent formulation is required for the effective demagnetizing field which allows to include the configuration dependent interaction field for an assembly of particles of arbitrary shape and outer volume and that allows determining how the interaction field will modify the demagnetizing field, the effective anisotropy or total energy and coercivity of the system.

The present study is focused in deriving mean field expressions for the configuration dependent effective demagnetizing and anisotropy fields in assemblies of exchange decoupled magnetic particles, which depend on both the shape of the particles, the volume containing them and the packing fraction. The magnetostatic properties of 2D and 1D assemblies have been analyzed and the results show that the dipolar interaction depends not only on the interparticle distance, but also on the outer shape of the assembly. In particular it is shown that this outer shape contribution to the dipolar field plays an important role in the coercivity of the assembly, the limiting values of the effective demagnetizing fields and the magnetization reorientation transition due to the dipolar interaction. Remanence curves and ferromagnetic resonance measurements done on arrays of bistable cylindrical nanowires have served first to identify the correspondence of these measurements with the model, and second, to validate this mean field approach.

\section{Experimental}
Arrays of Co, Ni$_{81}$Fe$_{19}$ and Co$_{55}$Fe$_{45}$ nanowires having a low density (low porosity $P$) and small diameter ($29 \geq \phi \geq 40$ nm) have been grown by electrodeposition into the pores of 21$\mu$m thick lab-made track-etched polycarbonate (PC) membranes, in which the pores are parallel to each other but randomly distributed.\cite{ferain}  Also, two samples have been fabricated using anodized allumina templates in order to have higher porosities (10 and 15\%). Full details of the preparation process can be found elsewhere.\cite{piraux2005}

\begin{figure}[t]
\begin{center}
\includegraphics[width=8.0cm]{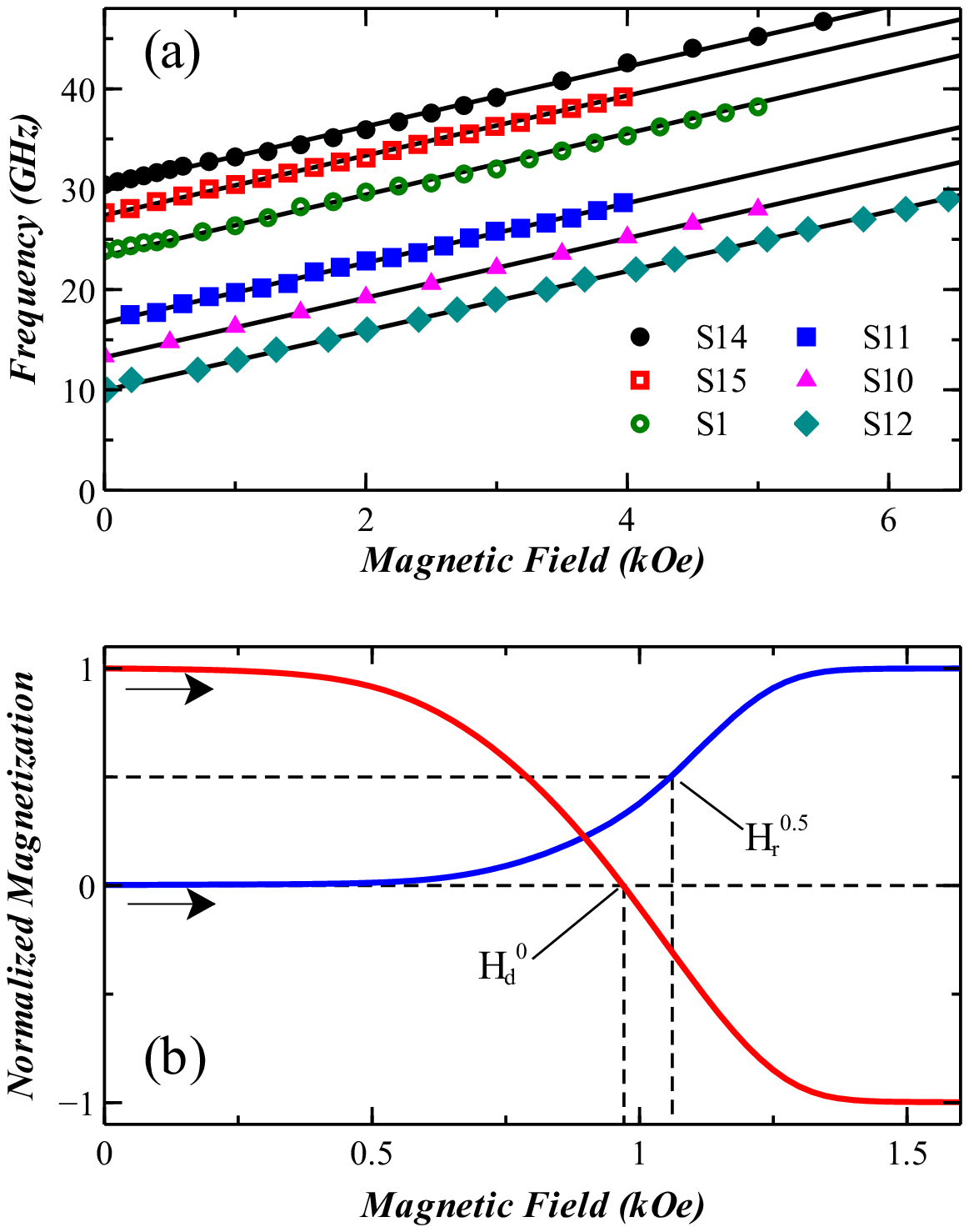}
\end{center}
\vskip -0.4cm
\caption{(a) Typical dispersion relations obtained on arrays of NWs with the field applied parallel to the wires, the measurements are shown as symbols, while the continuous line is the best fit using Eq. (\ref{reldisp3}). (b) IRM (blue) and DCD (red) remanence curves measured on sample S10 [NiFe] with the field applied parallel to the NWs axis.}
\label{disper}
\end{figure}

For the electrodeposition a Cr/Au layer is evaporated previously on one side of the membrane to serve as a cathode and deposition is done at a constant potential using a Ag/AgCl reference electrode. For CoFe a 40 g/l FeSO$_4$ + 80 g/l CoSO$_4$ + 30 g/l H$_3$BO$_3$ electrolyte was used with a potential of $V$=-0.9 V, while for NiFe the electrolyte contained 5.6 g/l FeSO$_4$ + 131.4 g/l NiSO$_4$ + 30 g/l H$_3$BO$_3$ and deposition is done at $V$=-1.1 V. 
Cobalt nanowires have been grown at $V$=-1V using a 238.5 g/l CoSO$_4$ + 30 g/l H$_3$BO$_3$ electrolyte with the pH set to 2.0 by addition of H$_2$SO$_4$ to favor a polycrystalline fcc-like Co structure with no magnetocrystalline anisotropy contribution.\cite{Darques} Table \ref{samples} shows the details of the samples considered.\\
\begin{table*}[ht] 
\caption{Co, NiFe and CoFe nanowire samples used in this study numbered as s1 to s16. For each sample the following quantities are given: the wire diameter $\Phi$, nominal packing fraction $P_N$, the FMR effective field $H_{eff}$, the dipolar interaction coefficient $\alpha_z$, and the determined values of the packing fraction $P_m$ and saturation magnetization $M^*_s$.}
\centering
\begin{tabular}{c c c c c c c c}
 \hline\hline
Material & Sample  & $\Phi (nm)$&$P_N$(\%) & $H_{eff}(kOe)$ &$\alpha_z$ (Oe) & $P_m$(\%) &$M^*_s$(emu/cm$^3$) \\ [0.5ex]
\hline
Co&s1&39&4.8 & 7.65 &400& 4.63& 1415 \\ \hline
Co&s2&40&5.6 &  7.48 &440& 5.09& 1406\\ \hline
Co&s3&30&3.9 &  7.57 &240& 2.91 & 1319\\ \hline
Co&s4&40&2.5 &  8.13 &140& 1.63& 1360\\ \hline
Co&s5&30&3.9 &  7.77&362& 4.08& 1409 \\  \hline
Co&s6&40&1.7 &  8.45 &118& 1.23 & 1396 \\ \hline
Co&s7&30&4.2 &  7.68 &370& 2.85 & 1337 \\  \hline
Co&s8&29&3.65 &  7.69 &350& 2.98& 1343 \\ \hline
NiFe&s9&30&3.46 &  4.64&140& 3.19& 816 \\ \hline
NiFe&s10&40&3.4 &  4.45 &180& 3.58& 792 \\ \hline
NiFe&s11&35&8.8 & 5.62 &580& 7.87 & 1171 \\  \hline
NiFe&s12&35&12 &   3.34&635& 11.52 & 812\\  \hline
CoFe&s13&40&3.9 &   10.52 &400& 3.42& 1868\\ \hline
CoFe&s14&29&5 &   10.24 &640& 5.23 & 1933 \\ \hline
CoFe&s15&29&5 & 9.18 &520& 4.77 & 1706 \\ \hline
CoFe&s16&18&10 &  8.45 &1274& 10.4& 1953 \\ 
\hline \end{tabular} 
\label{samples} 
\end{table*}
Ferromagnetic resonance (FMR) measurements have been done in the field swept mode with frequencies ranging from 1 up to 50 GHz using a 150 $\mu$m wide micro stripline with the DC magnetic field applied parallel to the long axis of the wires, as detailed elsewhere.\cite{piraux2005,Darques,Encinas} For all the samples considered, the transmission spectra is recorded at a fixed frequency while the magnetic field is swept from 10 down to 
0 kOe. These measurements are repeated for different frequencies and from the collection of frequencies and their respective resonance field, the dispersion relation is obtained. Figure \ref{disper} (a) shows typical dispersion relations obtained on the samples.
The dispersion relation for an array of infinitely long cylindrical nanowires with the field applied parallel to the wires which are in the saturated state is \cite{Encinas}
\begin{equation}
f = \gamma(H_{Res}+H_{eff}),
\label{reldisp3}
\end{equation}
where $H_{Res}$ and $H_{eff}$ are the resonance and effective anisotropy field, respectively. From the linear fit of the dispersion relation, see Fig. \ref{disper} (a); the values of  $H_{eff}$ have been determined for each sample, and the corresponding values are given in Table \ref{samples}.

Isothermal Remanence Magnetization (IRM) and DC demagnetization (DCD) remanence curves measurements with the field applied parallel to the wires axis, have also been done in all the samples, from which the interaction field coefficient along the wire axis, $\alpha_z$, has been determined as\cite{mtz}
\begin{equation}
\alpha_z = 2(H_r^{0.5}-H_d^0),
\label{alfa-exp}
\end{equation}
where $H_r^{0.5}$ is the field value at which the normalized IRM curve is 0.5, while $H_d^0$ is the field value at which the DCD curve is zero, as shown in figure \ref{disper} (b). The values determined for the interaction field coefficient $\alpha_z$ are given in table \ref{samples}.
\section{The effective demagnetizing field}
\begin{figure}[t]
\begin{center}
\includegraphics[width=8.0cm]{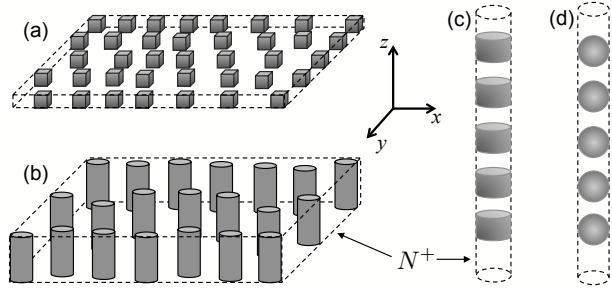}
\end{center}
\vskip -0.4cm
\caption{Schematics of an assembly of identical particles characterized by an inner demagnetizing factor $N$ contained within a macroscopic volume geometrically characterized by the outer demagnetizing factor $N^+$ (dashed lines).}
\label{system}
\end{figure}

Consider an assembly of exchange decoupled identical particles where each one has a demagnetizing factor given by $N$, called the inner demagnetizing factor, contained within a macroscopic volume described by an outer demagnetizing factor $N^+$, as schematically shown in figure \ref{system}. Furthermore, all the particles are aligned so their easy axis point in the same direction. The energy density of a given particle can include the magnetocrystalline ($K_{MC}$), shape ($K_{S}$) and magneto elastic ($K_{ME}$) anisotropies, which determine the total anisotropy of the particle, $K_A= K_{MC}+K_S+K_{ME}$, as well as a contribution due to the interparticle interaction ($E_{int}$)
\begin{equation}
E= K_{A}\sin^2\theta + E_{int}
\label{energy-gen}
\end{equation}
Where the shape anisotropy and the dipolar interaction are of magnetostatic origin and are related to the effective demagnetizing fields, which are considered in the following sections. Furthermore, in the following it will be assumed that the particles are homogeneously magnetized regardless of their shape.
\subsection{Saturated state}

The effective or total demagnetizing factor can be expressed using the interpolation procedure introduced by Netzelmann.\cite{netzelmann} To simplify the notation, in the following, the effective or total demagnetizing factor ($N_{eff}$) will be denoted as $N_T$ which can be expressed in terms of the packing fraction $P$ as,\cite{netzelmann,skomski} 
\begin{equation}
N_{T}=  (1-P)  N  +  P  N^+.
\label{nef-netzel}
\end{equation}
However, for the treatment of an assembly of magnetic particles it is more useful to rewrite this last expression as,
\begin{equation}
N_{T}=   N  +   (N^+-N)P.
\label{nef-us}
\end{equation}
The first term on the right side corresponds to the self demagnetizing factor of the particles that makeup the assembly, this is, $N_{self}$, while the second term contains all the contributions that depend on the packing and corresponds to the dipolar interaction contribution, $N_{dip}$, then
\begin{equation}
N_{T}=   N_{self}  +   N_{dip}.
\label{nef-us2}
\end{equation}
When $P= 0$ the demagnetizing factor reduces to that of the isolated non-interacting particle. While on the opposite limit, $P\rightarrow 1$ this leads to the outer demagnetizing factor. 
If both the inner and outer demagnetizing factors are diagonal and their trace is $\mathrm{Tr}(N)=4\pi$, then from Eq. (\ref{nef-us}), 
\begin{equation}
\mathrm{Tr}(N_{T})= \mathrm{Tr}(N) + [\mathrm{Tr}(N^+) - \mathrm{Tr}(N)] P,
\end{equation}
and from the form of (\ref{nef-us2}), the following properties of $N_T$ are obtained,
\begin{eqnarray}
\mathrm{Tr}(N_{T})&= &4\pi , \\
\mathrm{Tr}(N_{dip})&=& 0 .\label{dip1}
\end{eqnarray}
The $i$-th component of the effective demagnetizing field in the saturated state for an assembly of particles follows from equation (\ref{nef-us}),
\begin{equation}
H^{Di}_{T}=   M_s N_i  + (N_i^+-N_i)M_sP,
\label{hdef-us}
\end{equation}
where the first term is the usual self demagnetizing field of the particle, $H^D_{i}=M_sN_i$, and the second term is the dipolar interaction field, this is,
\begin{equation}
H^{i}_{dip}= (N_i^+-N_i)M_sP,
\label{hdef-us-pp}
\end{equation}
The total or effective magnetostatic anisotropy field in the saturated state defined as $H^S_T=M_s (N^x_T-N^z_T)$, where the hard axis is along the $x$-direction and the easy axis is taken along the $z$-axis, then,
\begin{equation}
H^S_T=M_s\Delta N + (\Delta N^+ - \Delta N)M_sP, 
\label{total-12field}
\end{equation}
The first term on the right side is the shape anisotropy field of the individual particles ($H_S$), while the second term corresponds to the total or effective dipolar field in the saturated state,
\begin{equation}
H_{dip}= (\Delta N^+ - \Delta N)M_sP, 
\label{total-12fieldx}
\end{equation}

The total shape (magnetostatic) anisotropy energy is given by $K^S_T = M_sH^T_S/2$, which includes both the shape anisotropy of the particle and the contribution of the dipolar interaction. This anisotropy constant now regroups $K_S$ and $E_{int}$ in equation (\ref{energy-gen}).

Finally, from the expressions obtained for the interaction field there are two other relations of interest that follow when both the inner and outer volumes have in plane symmetry, so that $N_x = N_y$. Due to the symmetry, the in-plane components of the interaction field are equal, $H^x_{dip}=H^y_{dip}$, and since Tr($N_{dip}$)=0, according to equation (\ref{dip1}), the following relation between the components of the dipolar fields is obtained,
\begin{equation}
H^{z}_{dip}= -2H^{x}_{dip}.
\label{2dipolar}
\end{equation}

This shows that when both the inner and outer volume have rotational symmetry, the dipolar interaction field along the symmetry axis is twice the dipolar field in the hard axis with opposite sign. This, for example, has been obtained by several authors.\cite{petrov,strijkers} A consequence of this result is that the dipolar part of the total or effective anisotropy field ($H_{dip}=H^x_{dip}-H^z_{dip}$) is
\begin{equation}
H_{dip}= - \frac{3}{2}H^{z}_{dip},
\label{dipo-l12}
\end{equation}
which provides a relation between the component of the interaction field along the easy axis, with the net contribution of the interaction field to the total anisotropy field.

\subsection{Non saturated states}
To extend the previous expressions to the non-saturated case, consider that the particles are bistable, this is, they can only be magnetized along their easy axis in both the positive or negative direction. The normalized magnetization, $m=M(H)/M_s$, so that $-1 \leq m \leq 1$. Any magnetic state called hereafter $m$, can be written in terms of the fraction of particles magnetized in the positive ($m_+$) and negative ($m_-$) directions, with $0\le m_{\pm}\le 1$, and $m=m_{+}- m_{-}$. Moreover, since the number of particles is constant, $m_{+}+m_{-}=1$, and $m_{\pm}$ can be obtained from the value of $m$ as,
\begin{equation}
m_{\pm}=\frac{1\pm m}{2}.
\label{m-plus}
\end{equation}
A recent FMR study done on non saturated arrays of bistable NWs lead to the expressions for the effective field, the dipolar interaction term and the FMR dispersion relation, that include explicitly the magnetic configuration of the system.\cite{intermag,joako-prb}  In particular, it was shown that to extend the known expressions for the saturated case to the non-saturated case, it is necessary to rewrite the dipolar interaction term as a parametric expression of $m_{\pm}$.\cite{intermag,joako-prb} Assuming a dipolar interaction field of the form $\alpha m$, it implies that the dependence of the interaction field on $m$ is not as a direct product but instead, requires using Eq. (\ref{m-plus}), this is,
\begin{equation}
H_{dip}m_{\pm}= \frac{H_{dip}}{2} \pm \frac{H_{dip}}{2} m,
\label{dipolar}
\end{equation} 
which at saturation reduces to the expected value. The $\pm$ sign is introduced to include both positive ($m=1$) and negative ($m=-1$) saturation.

Using Eq. (\ref{hdef-us}), the $i$th-component of the effective demagnetizing field is,
\begin{equation}
H^{Di}_{T}=M_sN_i+(N^+_i-N_i) \frac{M_sP}{2} \pm (N^+_i-N_i) \frac{M_sP}{2}m.
\label{demag-m}
\end{equation}
The dipolar contribution now contains two terms, the first one is constant, while the second one is proportional to $m$ and corresponds to the magnetization dependent part of the demagnetizing field. 
If the magnetization dependent interaction field is taken as $\alpha m$, then from the last term on the right hand side, the $i$-th component of the dipolar interaction field coefficient $\alpha$ can be identified as,
\begin{equation}
 \alpha_i = \frac{H^{i}_{dip}}{2} = (N^+_i-N_i) \frac{M_sP}{2}.
 \label{tk1}
\end{equation}
Then equation (\ref{demag-m}), can be written as, 
\begin{equation}
H^{Di}_{T}=H^D_i+\alpha_i \pm \alpha_i  m,
\label{tk2}
\end{equation}
where $H^D_i$ is the demagnetizing field of the isolated particle. From Eq. (\ref{total-12field}), the effective or total magnetostatic anisotropy field is, 
\begin{equation}
H^S_{T}=H_{S} + \alpha_T \pm \alpha_T m,
\label{total-heff}
\end{equation}
and $\alpha_T = \alpha_x - \alpha_z$ is the total dipolar field coefficient. 
At saturation $m=1$, both equations (\ref{tk2}) and (\ref{total-heff}) reduce to Eqs. (\ref{hdef-us}) and (\ref{total-12field}), respectively. 
\section{Applications}
\subsection{Two dimensional arrays with rotational symmetry}
An important class of magnetic assemblies are two dimensional arrays of exchange decoupled nanoparticles, ideally a monolayer of single domain particles, with perpendicular magnetization. These correspond, for example, to systems obtained by lithography, including perpendicular bit patterned media, also granular thin films with columnar structure as those used for perpendicular recording media, self assembled monolayers and nanowire arrays.\cite{terris2,lau}

For these systems one can assume that the height of the particles is very small compared with the lateral dimensions of the entire array and the volume containing the particles can be considered as an infinite thin film, so that $N^+_x = N^+_y = 0$ and $N^+_z = 4\pi$, as depicted in figure \ref{system} (a) and (b). Using these values in Eq. (\ref{hdef-us}) and assuming particles with in-plane symmetry so that $N_x =N_y$ and using $N_z=4\pi -2N_x$ to express quantities in terms of $N_x$, the  effective demagnetizing field perpendicular to the plane ($z$) is
\begin{equation}
H^{Dz}_{T} = 4\pi M_s  - 2N_x(1-P)M_s.
\label{3dipolar}
\end{equation}
Which shows that the effective demagnetizing field of a thin film made of exchange decoupled entities is less than $4\pi M_s$ by a quantity equal to $2N_x(1-P)M_s$.

The total shape anisotropy field [Eq. (\ref{total-heff})] is,
\begin{equation}
H^S_{T}= (3N_x-4\pi) M_s - \frac{3}{2}N_{x}M_{s}P \mp \frac{3}{2}N_{x}M_{s}Pm,
\label{dipo-qqq}
\end{equation}
the first term on the right side is the shape anisotropy of the particle, while the second and third term correspond to the dipolar part of the total anisotropy field, for $m=1$,
\begin{equation}
H_{dip}=- 3N_{x}M_{s}P,
\label{dipo-iii}
\end{equation}
which is a particular case of Eq. (\ref{dipo-l12}). As seen from this expression $H_{dip}$ is only a function of $N_x$ and $P$. Since the particles have in-plane symmetry, then $N_{z}=4\pi - 2N_x$, from where it follows that if $0 \leq N_z \leq 4\pi$, then $0 \leq N_x \leq 2\pi$. And from the previous expression for $H_{dip}$, it is possible to establish an upper bound for the dipolar part of the total or effective anisotropy. Taking $N_x = 2\pi$,
\begin{equation}
H^{max}_{dip}=- 6\pi M_{s}P.
\label{dipo-iv}
\end{equation}
The dipolar contribution to the total anisotropy field is antiferromagnetic, inferred by the negative sign in Eq. (\ref{dipo-iii}), as expected for a 2D array of nanomagnets with perpendicular anisotropy.

\subsection{Cylindrical Nanowires}
In the particular case of a 2D array of circular cylinders of arbitrary height aligned parallel to each other so their axes are along the $z$ axis, as shown in figure \ref{system} (b), their effective demagnetizing fields along the easy and hard directions are given by equation (\ref{hdef-us}). As in the previous section, the outer demagnetizing factor is taken as an infinite thin film. For a cylinder of arbitrary aspect ratio (wire height divided by its diameter), $N_{z}$ can be determined numerically.\cite{phatak} However, for the NWs considered in this study, the wire height is typically of the order of 20$\mu$m so the aspect ratio is very high, and they can be considered as infinite so their demagnetizing factors are $N_{z}=0$ and $N_{x}=N_y=2\pi$ and from equation (\ref{hdef-us}) the demagnetizing field along the easy and hard axis are,
\begin{eqnarray}
H^{Dz}_{T} &=& 4\pi M_{s}P, \label{wire-z}\\
H^{Dx}_{T} &=& 2\pi M_{s} - 2\pi M_{s}P. \label{wire-x}
\end{eqnarray}
Due to the symmetry properties of both inner and outer demagnetizing factors,  $H^{Dx}_{T} =H^{Dy}_{T}$ and the dipolar terms satisfy both equations (\ref{dip1}) and (\ref{2dipolar}). The effective anisotropy field is,
\begin{equation}
H^S_{T}= 2\pi M_{s}-6 \pi M_{s}P,
\label{heff} 
\end{equation}
that corresponds to the known expression for the effective field for an array of infinite cylindrical nanowires in the saturated state.\cite{Encinas} The effective dipolar interaction field is $H_{dip}= -6 \pi M_{s}P$, and using Eqs. (\ref{tk1}) and (\ref{total-heff}), the magnetization dependent part of the effective interaction field is,
\begin{equation}
\alpha_T = \frac{H_{dip}}{2} = 3\pi M_sP,
\label{atot-23}
\end{equation}
in agreement with the expression obtained in Ref. [\onlinecite{joako-prb}].

On the other hand, the measurement of the interaction field using Eq. (\ref{alfa-exp}) and the IRM and DCD remanence curves are done solely along the wire axis, this is, the measurement is related to the component of the interaction field along this direction, as pointed out in Ref. \onlinecite{mtz}. From Eq. (\ref{wire-z}), and Eqs. (\ref{tk1}) and (\ref{tk2}), the component of the magnetization dependent part of the  interaction field along the wire axis is
\begin{equation}
\alpha_z= 2\pi M_{s}P.
\label{alfa-z}
\end{equation}

Then, Eqs. (\ref{heff}), (\ref{atot-23}) and (\ref{alfa-z}) provide different expressions for the interaction field which have different physical meaning. Moreover, depending on the measuring technique, each of these three quantities can be measured independently,\cite{Encinas,joako-prb,mtz} so it is important to distinguish between them. 

\begin{figure}[t]
\begin{center}
\includegraphics[width=8.5cm]{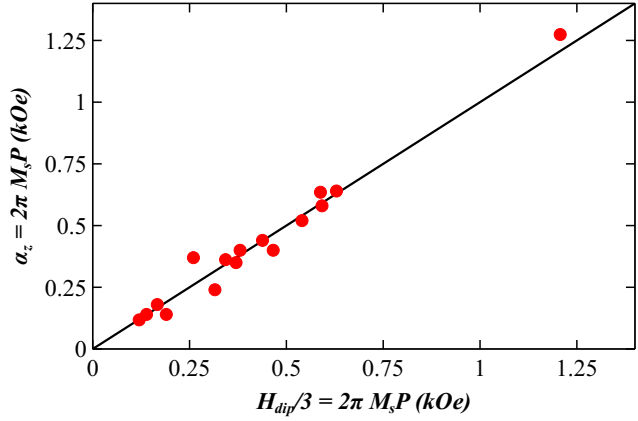}
\end{center}
\vskip -0.4cm
\caption{Easy axis component of the interaction field determined by magnetometry, $\alpha_z= 2\pi M_{s}P$, plotted as a function of $H_{dip}/3$, where $H_{dip}=6\pi M_sP$ is the total interaction field measured by FMR in the saturated state, for all the nanowire samples listed in Table \ref{samples}.}
\label{alfas}
\end{figure}

Since the effective anisotropy field, $H^S_{T}$ can be determined from the dispersion relation obtained from the FMR measurements, using equation (\ref{reldisp3}), then from Eq. (\ref{heff}),  $H_{dip}=H^S_{T}- 2\pi M_{s}$.  So $H_{dip}$ has been determined for each sample from the FMR measurements using the nominal values of $M_s$ for Co (1400 emu/cm$^3$), Ni$_{81}$Fe$_{19}$ (800 emu/cm$^3$), and Co$_{55}$Fe$_{45}$ (1900 emu/cm$^3$) and the values of $H_{eff}$ given in Table \ref{samples}. These values where then divided by 3 so they correspond numerically to Eq. (\ref{alfa-z}) which is the value of the component of the interaction field along the wire axis measured with the IRM and DCD remanence curves. Figure \ref{alfas} shows the values obtained using the IRM and DCD remanence curves, $\alpha_z$ plotted as a function of the values obtained by FMR, $H_{dip}/3$.

These results show a very good agreement, which shows that both Eqs. (\ref{heff}) and (\ref{alfa-z}) provide correct results for the interaction field.
Moreover, these two expressions provide expressions for quantities that can be measured independently, so its possible to find a self consistent method to determine both $M_s$ and $P$. Since $H^S_{T}$ and $\alpha_z$ are determined from FMR and magnetometry measurements, respectively, then combining Eqs. (\ref{heff}) and (\ref{alfa-z}) leads to the following expressions for $M_s$ and $P$.
\begin{equation}
M_{s} = \frac{H^S_{T}+3 \alpha_z}{2\pi},
\label{m-good}
\end{equation}
\begin{equation}
P =  \frac{\alpha_z}{H^S_{T}+3 \alpha_z}. 
\label{p-good}
\end{equation}

The values of $H^S_{T}$ and $\alpha_z$ measured by FMR and the remanence curves, respectively, that are given in Table \ref{samples} have been used as input in Eqs. (\ref{m-good}) and (\ref{p-good}) to obtain the corresponding values of  $M_s$ and $P$. The results are given in Table \ref{samples} in the columns labeled $P_m$ and $M^*_s$, respectively. As a first point, the values of $M^*_s$ show a very good agreement with the known values for Co, NiFe and CoFe. Furthermore the values for NiFe and CoFe alloys are in good agreement with those determined solely by FMR and reported elsewhere.\cite{joako-prb,joako-apl} Regarding the values of the template porosity, comparing the nominal values determined by scanning electron microscopy $P_N$ with those determined by the measurements $P_m$, a very good agreement is also found in practically all the samples. This procedure is general and can be extended to other assemblies by solving for $M_s$ and $P$ using Eqs. (\ref{total-12field}) and (\ref{tk1}). 

\subsection{Linear chain}
\begin{figure}[t]
\begin{center}
\includegraphics[width=8.5cm]{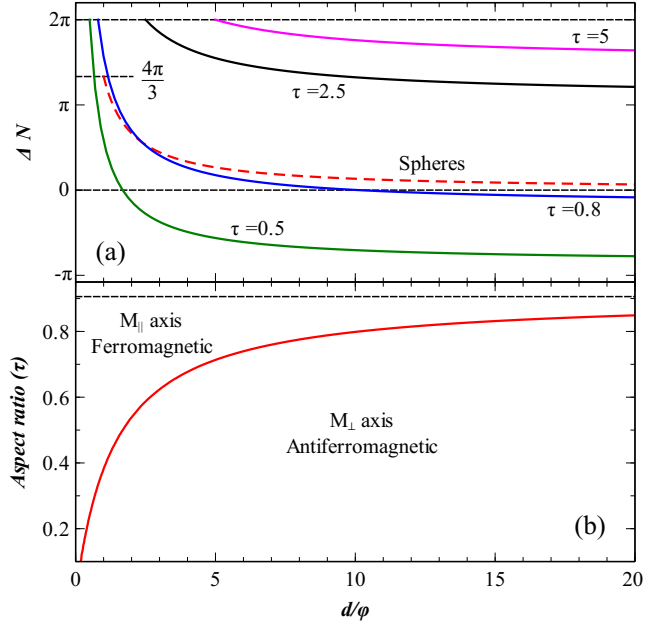}
\end{center}
\vskip -0.4cm
\caption{(a) Reduced effective anisotropy as a function of the reduced center to center distance $d/\phi$ for linear chains of cylinders with aspect ratio $\tau$=0.5, 0.8, 2.5 and 5 as well as for spheres (dashed line). (b) Zero anisotropy curve for a linear chain of cylindrical particles of aspect ratio $\tau$ as a function of the reduced center to center distance.}
\label{chains}
\end{figure}

So far, only 2D systems with perpendicular magnetization have been considered. Another interesting example is that of the linear chain of particles and particularly for circular cylinders of different aspect ratio and spheres, as those shown in figure \ref{system} (c) and (d), respectively. For an infinite number of particles, the outer demagnetizing factor is an infinite cylinder of circular cross section whose axis is assumed along the $z$ axis so $N^+_x=N^+_y=2\pi$ and $N^+_z=0$. While for both spheres and cylinders, the components of the inner demagnetizing factor are $N_x=N_y$ and $N_z=4\pi -2N_x$. From equation (\ref{total-12field}), the  effective anisotropy field is,
\begin{equation}
H^S_T= M_s(N_x-N_z) + M_s(6\pi-3N_x)P. \label{demag-h2}
\end{equation}
For the calculations, the demagnetizing factors and the packing fraction are required. For the chain of spheres, $N_x=N_y=N_z=4\pi/3$ and the packing fraction is $P=(2/3)\phi/d$ where $d$ is the center to center distance and $\phi$ is the diameter of the sphere. For the cylinders, $N_z$ is determined using known expressions as a function of the aspect ratio $\tau = h/\phi$ where $h$ is the height of the cylinder,\cite{phatak} and in this case, $P=h/d$.
Figure \ref{chains} (a) shows the reduced effective anisotropy $\Delta N= H^S_T/M_s$ for linear chains of cylinders with different aspect ratio ($\tau$=0.5, 0.8, 2.5 and 5) as well as for a chain of spheres (dashed line).

Consider first the case of the linear chain of cylinders. For a single, isolated and non interacting circular cylinder,  the critical aspect ratio is $\tau =0.9065$ above this value, $N_x > N_z$ and the easy axis is along the cylinder axis and at lower values $N_x < N_z$ and it is perpendicular to the axis. As seen in Fig. \ref{chains} (a) at large distances the anisotropy tends asymptotically to the expected value of the isolated non-interacting cylinder, which yield negative values for $\tau$ =0.5 and 0.8, and positive for  $\tau$ =2.5 and 5. 

As the particles are brought closer the interaction increases. For $\tau> 0.9065$, the interaction is ferromagnetic as it favors head to tail alignment of the magnetization, which results in an increase of the effective anisotropy as the packing fraction increases. For low aspect ratios ($\tau< 0.9065$) the easy axis of a given cylinder is perpendicular to the chain axis and the first term in Eq. (\ref{demag-h2}) is negative, so the interaction becomes antiferromagnetic as it has the opposite sign, and competes against it. As the packing fraction is increased, the value of the interaction increases and overcomes the anisotropy of the particle leading to a reorientation of the magnetization easy axis. 

In other words, to have a reorientation of the magnetization easy axis due to the dipolar interaction, the interaction has to be of the demagnetizing type (antiferromagnetic) with respect to the easy axis direction of the particle when considered isolated.  For the linear chain of cylinders there is a critical aspect ratio for the easy axis reversal which depends on the inter particle distance, that  correspond to those values for which the effective anisotropy vanishes. Setting $H^S_T =0$ in Eq. (\ref{demag-h2}) and using $2N_x=4\pi -N_z$, the critical packing fraction can be expressed as,
\begin{equation}
P_c=1-\left(\frac{4\pi}{3}\right)\frac{1}{N_z},
\label{critical-p}
\end{equation}
which is only valid for $N_z > 4\pi/3$. Here $N_z=N_z(\tau)$ gives the dependence on the aspect ratio of the cylinder which is given by well known expressions.\cite{phatak} 
Figure \ref{chains} (b) shows the critical aspect ratio variation with the reduced center to center distance for a linear chain of cylinders. Above this curve, the magnetization lies along the cylinder axis and the interaction between them is ferromagnetic. Below this line, the magnetization easy axis is perpendicular to the cylinder axis and the dipolar interaction is antiferromagnetic. At large distances, this curve tends asymptotically to $\tau = 0.9065$ (horizontal dashed line) which is the critical aspect ratio for an isolated non-interacting cylinder with $N_z = 4\pi/3$.\cite{phatak} The highest packing fraction required to reverse the easy axis is obtained when the axial demagnetizing factor of the disk $N_z$ tends to $4\pi$, which from Eq. (\ref{critical-p}), corresponds to $P_c=2/3$.

Finally, the linear chain of spheres presents features that are interesting as seen in Fig. \ref{chains} (a), in particular the existence  of an effective anisotropy and its increase as the packing fraction increases, which in light of  Eq. (\ref{demag-h2}) is entirely due to the dipolar interaction. Indeed, as seen in the figure, the effective anisotropy tends to zero as the distance between particles is increased consistent with a single sphere with no shape anisotropy. Moreover, the maximum value of the anisotropy is attained when the spheres come in contact and this value is lower than $2\pi$, which shows that there is not a full equivalence between a cylindrical wire and the chain of spheres. This point has been brought to light recently and its important since modeling of cylindrical wires is commonly done by considering them as a chain of spheres.\cite{bean,phatak2011} From Eq. (\ref{demag-h2}), the maximum value of the effective anisotropy for the chain of spheres is reached at $P=2/3$ and,
\begin{equation}
H^S_{Tmax}=\frac{4\pi}{3} M_s,
\end{equation} 
as indicated in Fig. \ref{chains} (a). 
\section{Discussion}
Expressions for the magnetization dependent demagnetizing field and the magnetostatic anisotropy field (energy) have been derived. These are general as the inner and outer volume are arbitrary and apply to assemblies of identical exchange-decoupled, bistable particles with a common easy axis. Moreover, this approach is a mean field approximation which only accounts for magnetostatic effects. In consequence, all the particles are equivalent which is only valid if the assembly contains a very large number of particles. 

Regarding the form of the effective demagnetizing field in the saturated state, Eq. (\ref{hdef-us}), it contains the self demagnetizing field of the individual particle and the component of the dipolar interaction field, Eq. (\ref{hdef-us-pp}) which is consistent with previous expressions obtained for nearly spherical particles.\cite{bookn}  On the other hand, when compared to other expressions,\cite{miles} which provide non physical results for particular limiting cases,\cite{richter1} Eq. (\ref{hdef-us}) provides correct limiting values when $P=$ 0 or 1. Moreover, based on more physical arguments and following Kronm\"{u}ller's analysis,\cite{kron} Drobrynin et al.,\cite{givord1} derived an expression for the effective demagnetizing field which contains a term that depends on the difference of the outer and inner demagnetizing factors as in Eq. (\ref{hdef-us-pp}). which provides support for Netzelmann's interpolation.\cite{netzelmann} 

The expression obtained for the effective demagnetizing field in the saturated state, Eq (\ref{hdef-us}), shows that this differs from that of the homogeneous material defined by the outer volume when $P<1$. For the particular case of a perpendicularly magnetized assembly with an outer volume of a thin film, as shown by Eq. (\ref{3dipolar}), the effective demagnetizing factor is lower than the continuous film, $N^z_T < 4\pi$ provided that $P<1$.\cite{seagate1,seagate2}

The extensions to include the configuration dependence are done by taking the product between the interaction term for the saturated case and Eq. (\ref{m-plus}) on the component of the demagnetizing field along the easy axis, Eq. (\ref{demag-m}), and on the total anisotropy field, Eq. (\ref{total-heff}).

An important result is that the interaction field can be expressed different forms each with different physical meaning, which however, are related among them. The dipolar interaction contributes to the components of the demagnetizing field, Eq (\ref{hdef-us}), as well as to the total magnetostatic anisotropy field, Eq. (\ref{total-12field}). Then each of these contributions can be extended to non saturated states assuming an interaction field of the form $\alpha m$, where the coefficient $\alpha$ is given by Eq. (\ref{tk1}). The FMR and magnetometry measurements done on arrays of nanowires provide a validation of this approach, while showing that different measuring techniques can provide values that correspond to different forms of expressing of the interaction field.

From the extension done to include the configuration dependency on the interaction field, the hysteresis loop shearing can be attributed to the magnetization dependent component of the effective demagnetizing field along the easy axis. As discussed in Ref. \onlinecite{mtz}, assuming an interaction field of the form $\alpha m$, the shearing can be corrected once the value of the coefficient is known, which; as shown here, is given by Eq. (\ref{tk1}).

The coercive field in an assembly is defined with respect to the total or effective anisotropy of the entire system, regardless of the specific internal composition, and in the Stoner - Wohlfarth model, these quantities are equal. If we call $H_c(0)=M_s\Delta N$ the coercive field of the isolated particle ($P=0$), rearranging Eq. (\ref{total-12field}), the coercive field of the assembly is given by,
\begin{equation}
H_c=H_c(0) (1-P) + M_s\Delta N^+ P,
\label{hc-p}
\end{equation}
which depends on the packing fraction as well as on the outer volume. The first term is the well known coercive field dependence on the packing fraction in assemblies where the shape anisotropy prevails. \cite{cullity,bate} The second term represents the outer volume anisotropy contribution, which implies that the coercivity of the assembly is shape dependent. So for a given assembly with fixed packing fraction, the coercive field will change between samples with different outer volume, as recently reported for granular hard magnetic films.\cite{givord3} Moreover, for the particular case when the outer volume is isotropic and $\Delta N^+=0$, the second term vanishes. To further emphasize the importance of the second term in Eq. (\ref{hc-p}), notice that if only the first term is considered, the coercivity always decreases when the packing fraction increases, regardless of how the particles are distributed.\cite{cullity,bate} This behavior is only consistent with the case where the interaction between the particles is antiferromagnetic. Indeed, for example, consider the linear chain of cylinders discussed above, each with their easy axis parallel to the chain axis, the interaction is ferromagnetic and as the packing fraction increases, the coercivity of the entire chain should increase from the value of the isolated cylinder of finite aspect ratio to that of the infinite cylinder. This behavior is described by Eq. (\ref{hc-p}) when the second term is taken into account. 

Since the total anisotropy field, Eq. (\ref{total-12field}), depends explicitly on this outer shape anisotropy $M_s\Delta N^+ P$, then variations on the shape of the outer volume results in changes in the demagnetizing field, the anisotropy and, as just mentioned, in the coercivity. This contribution is associated only to the dipolar interaction, so at low packing it is weak, while at higher packing fractions it can induce appreciable effects, and can be used as an additional parameter to control the magnetic properties of the assembly. An interesting case is when the assembly is made by spherical particles, since their shape anisotropy vanishes and the effective anisotropy field is given only by the outer shape anisotropy. Then a chain of spheres will behave close to a cylinder (its outer volume) and the entire chain has a uniaxial anisotropy whose origin is the interparticle interaction field,\cite{bean} while the same particles forming an spherical assembly will be isotropic. Although this is a well known result, it shows that this effect follows from the key role played by the outer demagnetizing factor on the value of the interaction field, which recently has been related to novel effects in granular hard magnets,\cite{givord3} composites,\cite{ramprasad} and dipolar ferromagnetic order in superparamagnetic particle monolayers.\cite{majetich} 

Another effect related to this contribution is the magnetization reorientation transition induced by the interaction field, which was discussed for the linear chain of cylinders, but which has also been reported for other assemblies.\cite{wires,Encinas,adeyeye11,tarta,tarta2} As discussed above, this transition results with increasing the packing fraction when the shape anisotropy of the single particle and the dipolar term have opposite signs. Qualitatively, this can also be interpreted as a competition between the shape anisotropies of the inner and outer volume and how their respective easy axis are oriented with respect to each other. At very low packing, the easy axis of a given particle in the assembly is that of the inner volume, while as $P \rightarrow$ 1 the easy axis corresponds to that of the outer volume. If both inner and outer volume have their easy axis in the same direction, there will be no reorientation transition of the magnetization, but if their easy axes are  not aligned, this transition will take place above a certain critical packing fraction. In this sense, both inner and outer volumes should be anisotropic, since if either one of them is isotropic, there will be no transition. Moreover, an anisotropic outer volume is a necessary condition to induce this transition as deduced from Eq. (\ref{hc-p}). Noting that in this expression $H_c$ is the total shape anisotropy and $H_c(0)=M_s\Delta N$ corresponds to the inner volume shape anisotropy, it follows that the total anisotropy will never change sign if the second term is not included.

The validity of this mean field approach is determined mainly by the extent to which the individual particles in the assembly remain bistable and are able to rotate independently from other particles, this is, that no collective effects take place while individual reverse their magnetic state. Starting with a very diluted assembly of bistable particles, both conditions are fulfilled, however as the packing fraction increases, each particle feels the interaction field which increases as the distance between particles is reduced. In this sense both bi-stability and the ability to reverse independently are susceptible to change as a consequence of any finite value of the interaction field. On one hand, bi-stability can be related to the strength of the coercive field of the individual particle, or to the height of the energy barrier that separates the two minima configurations ($+m$ and $-m$). On the other hand, the ability of the particles to rotate independently also depends on the value of the coercive field and the width of the intrinsic switching field distribution. So in both cases, the competition between the height of the intrinsic energy barrier of each particle and the magnitude of the interaction field determines if these conditions are fulfilled and in this sense, there is no unique criteria but rather it will depend on the type of material and the main contributions to the energy. 

In particular, the height of the energy barrier of the particles depends on the magnetocrystalline anisotropy. Since this contribution is intrinsic to each particle, it follows that the effects of the dipolar interaction are expected to become less relevant as the magnetocrystalline anisotropy of the individual particles increases.\cite{brown,schlomann1,schlomann2} 
In this sense, the magnetic hardness parameter\cite{coey-book} $\kappa=[|K_{MC}|/(2\pi M^2_s)]^{1/2}$, provides a useful empirical parameter of the height of the energy barrier. In particular, the higher the value of $\kappa$, the higher the packing fraction attainable without loss of bistability or independent reversal. On the other limit, this is, low values of $\kappa$ both the bi stability and independent reversal are expected to be limited to low and moderate packing fractions, as suggested by previous reports which show that magnetic percolation in assemblies takes place for packing fractions between 20 and 40\%.\cite{sousa,matei}

\section{Conclusion}
In conclusion, mean field expressions have been derived for the demagnetizing field and the magnetostatic anisotropy field which include the magnetization dependent part of the interaction field for assemblies of magnetic particles with arbitrary inner and outer demagnetizing factors. Special emphasis was given to 2D arrays of nanomagnets with perpendicular magnetization, where the expressions for the demagnetizing field and total anisotropy field have been derived. The model was successfully tested using FMR and magnetometry measurements done on arrays of cylindrical nanowires, where the equivalence between the interaction field determined by each of these techniques was shown. This formalism provides a simple mean field framework to describe magnetostatic effects in a wide variety of magnetic assemblies and composites. Moreover, it shows the role played by the outer demagnetizing factor in the value and characteristics of the interaction field which could be used to fine tune the magnetic properties of the assembly or composite.

\section{Acknowledgements}
\indent This work was partly supported by CONACYT M\'exico CB-105568, CB-177896  and 162651. J. M. Mart\'inez Huerta  and J. De La Torre Medina also thank CONACYT for financial supports No. 201754 and Retenci\'on-166089, respectively.

\end{document}